**Spectral analysis of the precessional switching of the magnetization in an isotropic thin film**

T. Devolder and C. Chappert.

Institut d'Electronique Fondamentale, UMR CNRS 8622, Bât. 220, université Paris-Sud, 91405 ORSAY, France.

_Abstract_ : *We study how a magnetic field step triggers the precessional switching of the magnetization in an isotropic thin film . Using a variational approach, we estimate analytically the switching frequency and compare it to the results obtained numerically by direct integration of the equations of motion. We show that the periodic motions of the three magnetization components can be described satisfactorily with truncated Fourier expansions, indicating a relatively high spectral purity of the magnetic response. Our analytical expressions are simple enough to be physically transparent at first sight, in contrast to the results of the more elaborate models that treat also anisotropy.*

Obtaining magnetization switching within the sub-nanosecond regime is currently one of the most challenged tasks in magnetism [1]-[5], both experimentally [4], [6] ,[7] numerically [5]-[8] and analytically [9]. In Magnetic Random Access Memories, it is of great industrial interest to design magnetization switching strategies that commute the storage elements magnetization in a fast way that can be easily scaled down.

When aiming at reversing the magnetization of a thin macrospin (i.e. a magnetic body with perfectly uniform magnetization) which is isotropic, the conventional strategy is to apply a field in the direction of the desired final magnetization state. This means often applying a field **H** antiparallel to the initial magnetization **M.** This generates no torque on **M** and is thus ineffective when fast reversal is willed.

When the macrospin has a uniaxial anisotropy, a more effective reversal strategy is a compromise between the torque magnitude and the imbalance of the energies of the two stable magnetization positions. In a famous paper [10], Stoner et al. showed that the best compromise was obtained for an angle of 135 degrees between **H** and **M**. Though cost-effective for the required applied field magnitude, the system needs to dissipate some energy to reach a stationary state which takes some nanosecond. When the field is switched off, it also takes a time delay for because the magnetization has relaxed to the easy axis.

In 1996, He et al. [11] studied the additional phenomena obtained when the field is applied in a fast-rising manner. Using numerical computations, they showed that due to the precessional nature of the magnetization motion, fast-rising field could induce very fast switching events for fields even smaller than the Stoner-Wohlfarth criterion. This was confirmed in more details by several authors [8], [9], [12]. The lowest switching field was predicted for an applied field perpendicular to the initial magnetization. The effect of finite rise-time and damping were studied [5], [8] by numerical integration of the Landau-Lifschitz equation (LL) [13]. Deep sub-ns precessional switching events, lasting half a precession period (typically 200 ps) were experimentally confirmed lately [2], [3].

The availability of analytical descriptions of precessional switching is highly desirable. This may first avoid time-consuming numerical integrations. Also it can help highlighting these when they can not be avoided. In recent Keystone papers, Serpico et al. [9] derived an exact analytical description of the precessional switching of a loss-free anisotropic macrospin. The switching time was expressed by means of Jacobi elliptic functions and indefinite integrals, while the magnetization trajectory was expressed using rational fractions of Jacobi ellictic functions. Although this description is quite powerful, the physical parameters that are relevant for the switching speed do not come out straightforwardly from their formalism.

In this letter, we show that if in addition to ref. [9] the macrospin has an isotropic plane, the frequency and the magnetization time-response have quite simple expressions, from which the relevant physical parameters are clear. It is found that the switching frequency scales with $\sqrt{HM_S}$, the square root expressing why the reversal is much faster than the magnitude of applied fields could make us expect. Using a variational approach, we also derive the time response of magnetization by quickly converging trigonometric expansions. The periodic motion of the magnetization has a surprisingly high spectral purity, which could not be ascribed from previous studies incorporating finite anisotropy [9].

## 1. <u>Geometry and notations</u>

In the absence of losses, the dynamics of a macrospin can be described by the Landau-Lifshitz equation [13]:

$$\text{Eq. 1} \qquad \mathbf{\dot{M}} = \gamma_0 \, \mathbf{H}_{\text{eff}} \times \mathbf{M}$$

where the point denotes the time derivative. We write $\gamma_0 = \gamma\mu_0 > 0$ and $\gamma/2\pi = 28 \, GHz/T$ the gyromagnetic factor. The effective field is the sum of the applied field $\mathbf{H}$ and the demagnetizing field $\mathbf{H}_D$. Throughout this paper, international SI units are used : $\mathbf{H}$ and $\mathbf{M}$ are in A/m, $\mu_0 H$ and $\mu_0 M_S$ are in Tesla, and $\omega_H = \gamma_0 H$ and $\omega_M = \gamma_0 M_S$ are frequencies.

We assume an isotropic thin film with infinite lateral extension in the (xy) plane, such that the demagnetizing field is solely along (z) and is $H_{demag} = -m_z M_S$, where $\mathbf{m} = \mathbf{M}/M_S$.

The field $\mathbf{H}$ is applied along the +(y) direction, i.e. perpendicularly to the initial magnetization which is assumed along +(x). We define $h = H/M_S > 0$ its reduced strength. Exact expressions where we assume $h << 1$ are displayed using the symbol "$\cong$".

Rewriting the Landau-Lifshitz equation in (xyz) coordinates yields:

Eq. 2 $$\dot{m}_x = \omega_M m_z (m_y + h)$$

Eq. 3 $$\dot{m}_y = -\omega_M m_z m_x$$

Eq. 4 $$\dot{m}_z = -\omega_M h m_x$$

The next section (§2) aims at describing the magnetization trajectories by solving the previous 3 equations. These trajectories could equivalently be derived from the conservation of the magnetization norm $m^2=1$ and of the energy $E_{tot}$:

Eq. 5 $$E_{tot} = -\mu_0 M_S^2 h m_y - \frac{1}{2}\mu_0 M_S^2 m_z^2$$

However, we will proceed using the previous LL equations so as to account also for the rate of change of the magnetization.

## 2. Precessional trajectory for a loss-free isotropic thin film

Using the initial conditions $m_x=1$ and applying abruptly the field $h$ at $t=0$, the magnetization trajectory in the *(yz)* plane can be obtained using Eq. 4 to eliminate $m_x$ in Eq. 3:

Eq. 6 $$m_z^2 = 2h m_y$$

It follows that $m_y$ obeys $m_y \geq 0$. The y-component of the magnetization always points towards the applied field. Inserting Eq. 6 in Eq. 2 and multiplying by Eq. 4 gives the trajectory in the *(xz)* plane:

Eq. 7 $$m_x^2 = 1 - m_z^2 - \frac{m_z^4}{4h^2}$$

Or equivalently in the *(xy)* plane:

$$\text{Eq. 8} \qquad m_x{}^2 = 1 - 2hm_y - m_y{}^2$$

These trajectories are displayed in Figure 1 for a typical value of *h=0.01*, which would correspond to 10.8 mT (108 G) for a permalloy macrospin. The **M** trajectory is almost semi-circular (Figure 1A) in the (xy) plane and does not excurse much in the (xz) plane (Figure 1B). From Eq. 7 and Eq. 8, we get the maxima of $m_z$ and $m_y$ occuring for *mx=0*:

$$\text{Eq. 9} \qquad m_z{}^{max} = \pm\sqrt{2h\left(\sqrt{1+h^2}-h\right)} \cong \pm\sqrt{2h}$$

$$\text{Eq. 10} \qquad m_y{}^{max} = \sqrt{1+h^2} - h \cong 1 - h$$

In practical cases where *h<<1*, we have $H_{demag}{}^{max} \cong -\sqrt{2HM_S}$ . The square root can be seen as an amplification factor: the maximum demagnetizing field is much greater than the applied field. For instance, if $\mu_0 H = 10.8$ mT, then $\mu_0 H_{demag}$ can reach up to 152 mT in Permalloy. This corresponds to a maximum out-of-plane incursion angle of 8 degrees. At this point, the magnetization rotates very fast and its rate of change is $\dot{m}_x \cong \gamma_0\sqrt{2HM_S}$

## 3. Time evolution of the magnetization

Along the trajectory, the energy flows back and forth from the Zeeman contribution when $|m_x|=1$ to the demagnetizing energy when *mx=0*. Hence, the effective field magnitude varies a lot along the trajectory (see Table 1) and can not be used to guess a characteristic frequency. Hence, there is a need for accurate analytical estimate of the magnetization time response. We first (§3.1) derive the main trends using basic trial functions, and then expand **m(t)** in a Fourier development.

### 3.1. Trigonometric trial functions and rough estimate of the main frequency

From Table 1 and Figure 1, intuitive trial functions for the time dependence of magnetization are:

$$\text{Eq. 11} \qquad m_x(t) = \cos \omega_r t$$

$$\text{Eq. 12} \qquad m_y(t) = m_y^{\ max}\left(1 - \cos 2\omega_r t\right)/2$$

$$\text{Eq. 13} \qquad m_z(t) = -m_z^{\ max} \sin \omega_r t$$

The $m_y^{\ max}$ and $m_z^{\ max}$ are those derived formerly (Eq. 9, Eq. 10). The "$r$" subscript to the angular frequency $\omega_r$ is used to recall that these trial functions are rough estimates. They do satisfy the magnetization norm conservation at the points listed in Table 1. Elsewhere, the norm oscillates typically between 0.8 and 1.

Using the classical properties of the trigonometric functions, Eq. 13 and Eq. 4, we get:

$$\text{Eq. 14} \qquad \omega_r = \omega_H / m_z^{\ max} \cong \gamma_0 \sqrt{HM_S/2}$$

The same result in obtained by inserting Eq. 12 in Eq. 3. From the above estimates, it is likely that the frequency of the magnetization evolution scales with $\gamma_0 \sqrt{HM_S}$.

Note that this could also be derived by combining Eq. 2 and Eq. 4 to achieve a second order differential equation of $m_z$:

$$\text{Eq. 15} \qquad \ddot{m}_z + \omega_M^{\ 2} h\left(h + m_y\right) m_z = 0$$

This equation is harmonic when $m_y = m_y^{max} \approx 1$ because $\dot{m}_y = 0$. It also leads to a characteristic frequency scaling with $\gamma_0 \sqrt{HM_S}$. Obtaining the scaling factor requires to develop $\boldsymbol{m(t)}$ in its main Fourier components.

### 3.2. Fourier expansions of the trial functions

From Eq. 4, $m_z$ and $m_x$ have the same fundamental frequency $\omega$. From Eq. 6, the fundamental frequency of $m_y$ is $2\omega$, and its development thus contains only harmonics of even order (i.e $2\omega$, $4\omega$…). From this and Eq. 6 we can also deduce that $m_z$, hence $m_x$ (see Eq. 4) contains only harmonics of odd order. Finally, since $\dot{m}_x(t=0) = 0$ and $\dot{m}_y(t=0) = 0$, the developments of $m_x$ and $m_y$ shall contain only cosine terms. From Eq. 15, we have $\ddot{m}_z(t=0) = 0$ such that the $m_z$ development contains only sine terms.

Let us continue our variational approach and consider improved trial functions:

$$\text{Eq. 16} \qquad m_x(t) = X_1 \cos \omega_r t + X_3 \cos 3\omega_r t$$

$$\text{Eq. 17} \qquad m_y(t) = Y_0 - Y_2 \cos 2\omega_r t - Y_4 \cos 4\omega_r t$$

$$\text{Eq. 18} \qquad m_z(t) = -Z_1 \sin \omega_r t - Z_3 \sin 3\omega_r t$$

These improved trial functions can be used to gain a better accuracy in estimating the solutions of the LL equation. We have 8 unknown parameters: the 7 Fourier components and the frequency $\omega$. Each of the 3 LL equations contains two harmonics, which provides thus 6 equations. The initial conditions give the needed further two equations: $m_x(t=0)=1$ provides $X_1+X_3=1$ and $m_y(t=0)=0$ provides $Y_0-Y_2-Y_4=0$.

The solving procedure is reported here below. The assumptions are that $Z_3 << Z_1$, $X_3 << 1$ (which were checked a posteriori) and that $h << 1$.

Eq. 4 applied on the improved trial functions leads to:

Eq. 19 $$\omega / \omega_H = X_1 / Z_1 = X_3 / 3Z_3$$

Eq. 3 leads to the evaluation of $Y_2$ and $Y_4$.

Eq. 20 $$Y_2 = \frac{\omega_M}{4\omega} \left( X_1 Z_1 + X_1 Z_3 - X_3 Z_1 \right) = \frac{\omega_M \omega_H}{2\omega^2} \left( X_1^{\,2} / 2 - X_1 X_3 / 3 \right)$$

Eq. 21 $$Y_4 = \frac{\omega_M}{4\omega} \left( X_1 Z_3 + X_3 Z_1 \right) = \frac{\omega_M \omega_H}{3\omega^2} X_1 X_3$$

Eq. 2 provides another relationship between all terms:

Eq. 22 $$\omega X_1 = \omega_H Z_1 + \omega_M \left( Z_1 Y_0 + Z_1 Y_2 / 2 - Z_3 Y_2 / 2 \right)$$

Using $h<<1$ and assuming $Z_3<<Z_1$, we get by using Eq. 19:

Eq. 23 $$\omega_H X_1^{\,2} \approx \omega_M Z_1^{\,2} \left( Y_0 + Y_2 / 2 \right)$$

Using Eq. 19, and $Y_0 \approx Y_2 \approx 0.5$, we get an improved [14] estimate of the main frequency:

Eq. 24 $$\omega = \gamma_0 \sqrt{3HM_S / 4}$$

Let us now estimate the values of all Fourier components.
Using $X_1 + X_3 \approx 1$ and assuming $X_3 << 1$, Eq. 20 yields:

Eq. 25 $$X_3 \approx \frac{3}{8} - \frac{3\omega^2 Y_2}{2\omega_M \omega_H}$$

The above parametric equations must be solved in a self-consistent way. Starting from $X_1=1$, $Y_2=0.5m_y^{max}$, $Z_1=m_z^{max}$ and Eq. 24, two recursive uses of Eq. 23 and Eq. 25 lead to partial convergence [15] to the approximate solutions:

Eq. 26 $\qquad X_1 \approx 9/8, \quad X_3 \approx 1/8, \; Y_0 \approx (29/64)m_y^{max}, \quad Y_2 = (1/2)m_y^{max}, \quad Y_4 \approx (3/64)m_y^{max},$

$$Z_1 = 23m_z^{max}/24, \quad Z_3 = - m_z^{max}/24.$$

### 3.3. Discussion

The improved frequency estimate in Eq. 24 may be compared to the periodicity that can be obtained by direct numerical integration [16] of the LL equation. Such a comparison, displayed in Figure 2A, indicates that the analytical frequency in Eq. 24 overestimates the real frequency by typically 3 %. The frequency obtained through numerical integration obeys the empirical law:

Eq. 27 $\qquad \omega = \gamma_0 \sqrt{0.71 H M_S}$

Which falls very near our estimate Eq. 24.

It is worth recalling that the small angle magnetization eigen excitations of the system are well known from the ferromagnetic resonance theory [17]. During a Uniform Precession (UP), the magnetization rotates around the applied field with a frequency $\omega_{UP} = \gamma_0 \sqrt{H(H+M_S)} \cong \gamma_0 \sqrt{H.M_S}$. Although our trajectories very significantly depart from small angle excitations, it is remarkable that the characteristic frequencies scale with the same quantity. The main difference comes from the existence of a substantial variation of $m_y$ and from the spectral weight of the harmonics.

The first order (§3.1) and higher order (§3.2) Fourier developments are compared to the numerical integration of LL for $h=0.01$ in Figure 2B. The time coordinate has been normalized to the respective precession frequencies to ease the comparison. The first order expansion clearly fails at giving the right time evolution of $m_x$ because of a significant $3\omega$ harmonics, while the evolution of $m_y$ and $m_z$ is satisfactory accounted for.

The agreement is almost perfect if the $3\omega$ and $4\omega$ contributions are considered (Figure 2B). With no fitting parameter and the spectral weights of Eq. 26, the overall maximum absolute disagreement is 3% in the magnetization components. The magnetization norm (not shown) fluctuates between 0.969 and 1.

To summarize, the analytical expressions describing the magnetization motion of a macrospin-like thin film with neither anisotropy nor damping can account with 3% accuracy for the switching frequency, and with 3% accuracy for the frequency spectrum of the magnetization response.

## 4. Conclusion

We have studied analytically the precessional switching of an isotropic thin film macrospin subjected to field step in the in-plane direction perpendicular to the initial magnetization. During the reversal, the precession pulls the magnetization out of the film plane, thus creating a significant demagnetizing field. It reaches a magnitude $H_{demag}^{\ max} \cong -\sqrt{2HM_S}$ much higher than the applied field, which significantly accelerates the magnetization rotation, allowing deep sub-ns switching times. The correlated characteristic switching frequency is $\omega = \gamma_0 \sqrt{0.71HM_S}$. The periodic motions of the three magnetization components have relatively high spectral purities. The first two non-vanishing terms in the Fourier expansion Eq. 26 are sufficient to describe the time-resolved response of the magnetization with 3% accuracy. The $m_x$ component looks mainly like a cosine function, whereas the $m_y$ component looks like a cosine at double frequency.

**Tables**

| Point of evaluation | $m_x$ | $m_y$ | $m_z$ | $\dot{m}_x$ | $\dot{m}_y$ | $\dot{m}_z$ | Comment |
|---|---|---|---|---|---|---|---|
| **m** above the +(y) axis | 0 | $\cong 1-h$ | $\cong \pm\sqrt{2h}$ | $\cong \gamma_0 \sqrt{2HM_S}$ | 0 | 0 | $\ddot{m}_x = 0$ |
| **m** along the $\pm$(x) axis | $\pm 1$ | 0 | 0 | 0 | 0 | $\mp \gamma_0 H$ | $\ddot{m}_z = 0$ |

Table 1: coordinates of some particular points of the trajectory of the magnetization vector of an isotropic thin macrospin.

**Figure captions**

Figure 1 : Exact trajectory of the magnetization vector a loss-free macrospin film of initial magnetization along (x) when subjected to a transverse field *h=0.01* applied along (y).

Up inset: quasi half-circle trajectory in the (xy) plane (easy plane).

Down inset: trajectory projected in the (xz) plane (out of the easy plane incursion).

Figure 2 : **A:** Frequencies ($\omega/2\pi$) obtained through numerical integration of LL using OOMMF software compared to the analytical estimate of Eq. 24.

**B**: Comparison between the time evolution of the magnetization obtained through numerical integral of LL (symbols) with the Fourier serie expansions. The first order (up to *2ω*) and the second order (up to *4ω*) are displayed respectively with dashed-dotted bold line, and (resp.) with the thin continuous lines very near the symbols.

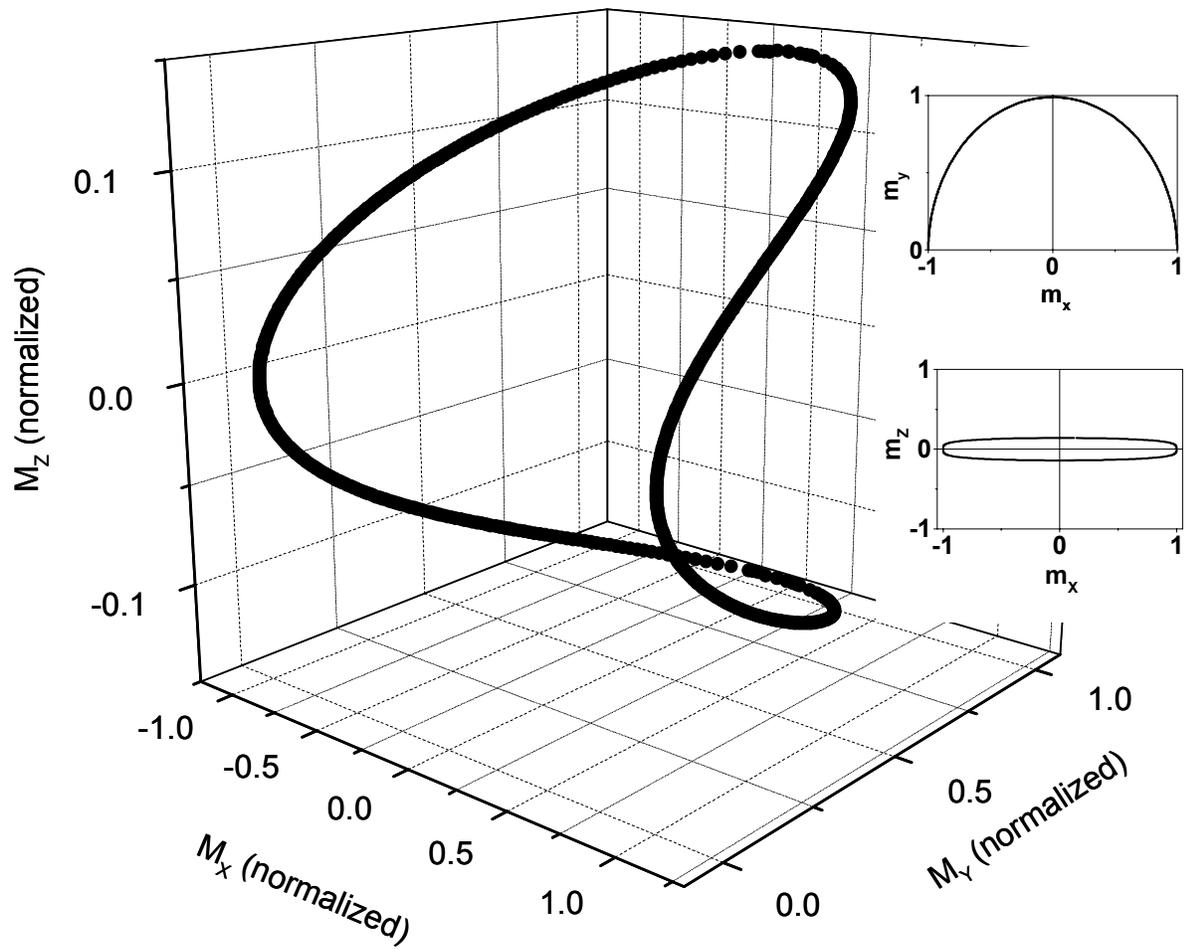

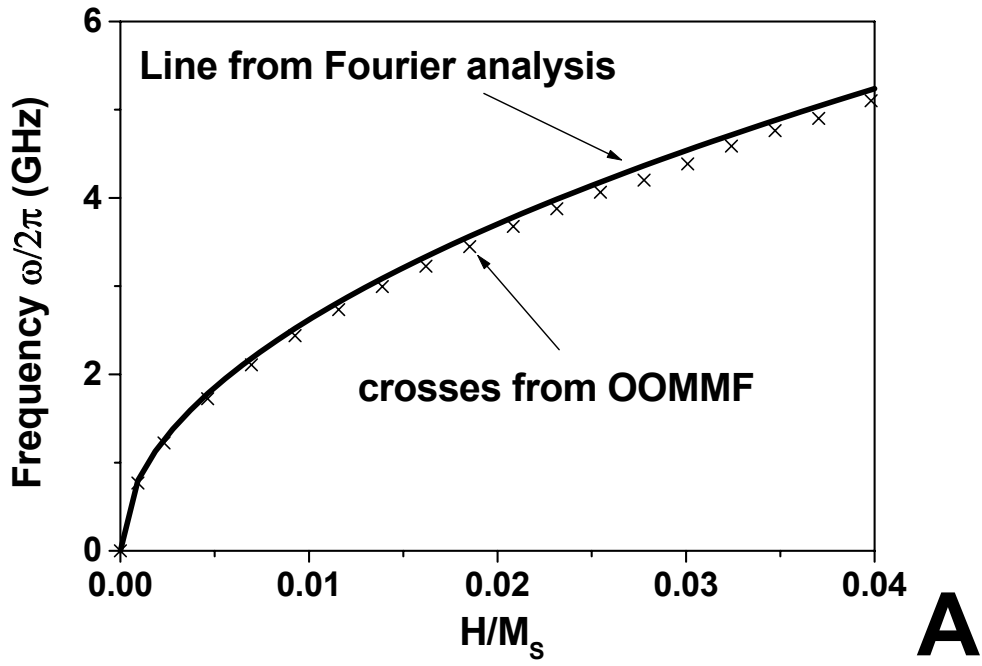

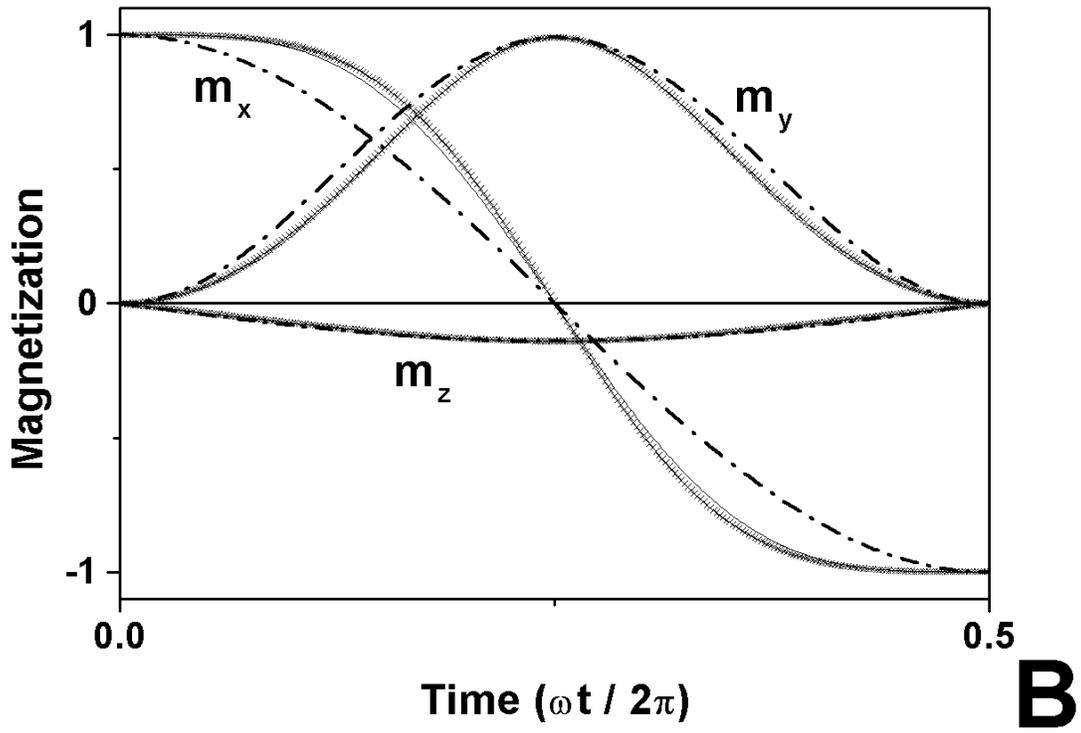